\documentclass[12pt,a4paper]{article}

\DeclareFontFamily{OT1}{rsfs}{}
\DeclareFontShape{OT1}{rsfs}{m}{n}{ <-7> rsfs5 <7-10> rsfs7 <10->
rsfs10}{} \DeclareMathAlphabet{\mycal}{OT1}{rsfs}{m}{n}

\newcommand{\I}{\mycal I^+}

\newcommand{\be}{\begin{equation}}
\newcommand{\ee}{\end{equation}}
\newcommand{\beq}{\begin{eqnarray}}
\newcommand{\eeq}{\end{eqnarray}}

\usepackage{amsmath,amsthm,amsfonts}
\input amssym.def
\newsymbol\square 1003

\begin{document}
\centerline{\bf  LAGRANGIAN AND HAMILTONIAN FOR THE BONDI-SACHS METRICS}

\bigskip

\centerline{J. Korbicz* and J. Tafel**}

\noindent
*Institute of Theoretical Physics, University of Hannover, Appelstr. 2, D-30167 Hannover, Germany, email: jarek@itp.uni-hannover.de\\
**Institute of Theoretical Physics, University of Warsaw,
Ho\.za 69, 00-681 Warsaw, Poland, email: tafel@fuw.edu.pl

\bigskip

\noindent
{\bf Abstract}. We calculate the Hilbert action for  the Bondi-Sachs metrics. It yields  the Einstein vacuum equations in a closed form. Following the Dirac approach to constrained systems we investigate the related Hamiltonian formulation. For regular null surfaces the Hamiltonian defines an energy which coincides with the Bondi mass and is positive.

\bigskip


\null

\section{Introduction}

This paper is a follow up of Refs \cite{TP}\cite{T} in which we studied the Bondi-Sachs metrics \cite{B}\cite{S} in relation to the Penrose conformal approach to the asymptotic flatness \cite{P}\cite{PR}.
Here we consider the Lagrangian and Hamiltonian formulation of the Einstein theory (in vacuum) for these metrics. This formalism is related to a null slicing of spacetime. We derive the exact Einstein equations and we single out those of them which are constraints from the point of view of an evolution with respect to the Bondi time $u$. We perform also a preliminary study of the related canonical formulation in a manner similar to that of Goldberg \cite{G}\cite{G1} (see also  \cite{GR}, where the Ashtekar variables are used for a null foliation). Our approach differs from that of Torre \cite{TO} and from the canonical formalism on asymptotically null hyperboloids (see e.g. \cite{AS} \cite{J} and references therein). In the last section we investigate relations between the Hamiltonian and the Bondi mass (also called Trautman-Bondi energy).

Unlike in \cite{TP}\cite{T}, in this paper   we consider Bondi-Sachs metrics using the affine distance  instead of the luminosity one. Thus, we assume that   spacetime metric $\tilde g$ of signature $+---$ takes the form
\begin{equation}
\tilde g=du(\tilde g_{00}du+2dr+2\tilde \omega_{A} dx^A)+r^2g_{AB}dx^Adx^B\ .\label{3}
\end{equation}
The coordinates $u$ and $r$ are interpreted, respectively,  as a retarded time and a distance from matter sources and $x^A$, $A=2,3$, are coordinates on the 2-dimensional sphere $S_2$ with the standard metric $s_{AB}$. 
The function $\tilde g_{00}$, the form $\tilde \omega_{A} dx^A$ and the tensor $g_{AB}dx^Adx^B$ are regular on  $[a,b]\times [r_0,\infty]\times S_2$. 
In order to guarantee the asymptotic flatness of $\tilde g$ at null infinity these objects are assumed to admit the following expansions with respect to $\Omega=r^{-1}$
\begin{equation}
\tilde g_{00}=1-2M\Omega+O(\Omega^{2})\ ,\label{4}
\end{equation}
\begin{equation}
\tilde \omega_{A}=\psi_A+\Omega\kappa _A +O(\Omega^{2})\ ,\label{6}
\end{equation}
\begin{equation}
 g_{AB}=-s_{AB}+\Omega n_{AB}+O(\Omega^{2})\ ,\label{7}
\end{equation}
with coefficients depending on  $u$ and $x^A$. 
In accordance with the Einstein equations  we also assume that
\begin{equation}
s^{AB}n_{AB}=0\ .\label{8}
\end{equation}
Above  properties of $\tilde g$ assure that the conformally equivalent metric 
$g=\Omega^2\tilde g$
is well defined up to the scri $\I$ \cite{PR}, where $\Omega=0$. 

Equation (\ref{8}) is a reminiscent of the luminosity condition
\begin{equation}
\det g_{AB}=\det s_{AB}\label{lum}
\end{equation}
considered in original papers of Bondi et al \cite{B} and Sachs \cite{S}.  In this paper we replace (\ref{lum}) by (\ref{8}) and 
\be
\tilde g_{01}=1\label{12}
\ee
in order to simplify the Lagrangian formulation.  Now $r$ is the affine parameter for null geodesics generating surfaces $u=$const, 
\be
\tilde g^{\alpha\beta} u_{|\alpha}\partial_{\beta}=\partial_r\ ,\label{14}
\ee
but $M$ differs from the Bondi mass aspect. 
Under assumptions (\ref{3})-(\ref{8}) one can always pass from (\ref{12}) to (\ref{lum}) by means of the transformation
\be
r=ar'\ ,\label{9a}
\ee
where 
\be
a=({\det s_{AB}\over \det g_{AB}})^{1/4}=1+b\Omega^{2}+O(\Omega^{3})\label{9b}
\ee
and $b$ does not depend on $r$.
Under this transformation the fields $n_{AB}$  and $\psi_A$ do not change and 
\be
M'=M-b_{,0}\ ,\label{15}
\ee
\be
\kappa'_A=\kappa_A+b_{,A}\ .\label{17}
\ee 
 Note that $M'$ is the Bondi mass aspect.

Substituting (\ref{3})-(\ref{8}) into the Einstein vacuum equations yields 
\begin{equation}
\psi_A=-{1\over 2}n^B_{\ A|B}\ ,\label{9}
\end{equation}
\be
g_{AB}=-(1+\frac{1}{8}\Omega^2 n^2)s_{AB}+\Omega n_{AB}+O(\Omega^3)
\label{17b}
\ee
(hence $b=\frac{1}{16}n^2$) and the following relations between 
  the  u-derivatives, denoted by a dot, of the coefficients $M$, $n_{AB}$ and  $\kappa_A$
\be
\dot {\hat M}=-{1\over 8}\dot n^{AB}\dot n_{AB}\ ,\label{27}
\ee
\be
\dot \kappa_A=(\frac{1}{4}n\dot n-\frac{2}{3}M)_{,A}-\frac{1}{3}n_{AB}\dot 
n^{BC}_{\ \ \ |C}+(\frac{1}{6}\dot n_{AC}n^C_{\ B}-\frac{1}{3}n^C_{\ [A|B]C})^{|B}\ .
\label{28}
\ee
 Here
\be
\hat M=M-\frac{1}{8}n\dot n+{1\over 4}n^{AB}_{\ \ \ |AB}\ ,\label{17d}
\ee
$n^2$ denotes $n_{AB}n^{AB}$ and the metric $-s_{AB}$  is used to lower and raise the indices $A,B$ and to 
define  the covariant derivatives $_{|A}$ (note that this convention changes in  the next sections).

Concerning a physical meaning of different fields on $\I$ we note that $\hat M$ can be used instead of the Bondi mass aspect to describe the total energy 
related to a section $\Sigma$ of the scri $\I$  \cite{C}\cite{T}
\be
E(\Sigma)=\frac{1}{4\pi}\int_{\Sigma}\hat M d\sigma\ .\label{10a}
\ee
Equation (\ref{27}) is responsible for  the energy loss formula \cite{TR}\cite{B}\cite{S}. The traceless tensor $\dot n_{AB}$ corresponds to the Bondi news function.  Functions $\kappa_A$ play an important role in a description of the angular momentum at null infinity (see e.g. \cite{PR} and references therein).

\section{Lagrangian and equations}

\null

In this section we consider the Hilbert action for metrics (\ref{3})
\be 
-\gamma\int d^4x\sqrt{|\tilde g|}\tilde R\ , \label{26}
\ee
where $\gamma=1/16\pi$ in the units $c=G=1$.  Since  the metric components 
$\tilde g_{1\mu}$ are fixed the variational principle for (\ref{26})
yields all the Einstein equations except $\tilde G^{1\mu}=0$, where $\tilde 
G^{\mu\nu}$ is the Einstein tensor of $\tilde g$. The lacking equations  are also fulfilled  provided all other Einstein equations are satisfied as well as asymptotic conditions (\ref{4})-(\ref{8}) and  equations (\ref{27}), (\ref{28}) (see \cite{S} \cite{NU} and  discussion at the end of this section). The boundary equations (\ref{27}), (\ref{28}) require a separate treatment. For instance, one can obtain them from a Lagrangian defined on $\I$ with a help of Lagrange multipliers. Another possibility is to treat boundary equations (\ref{27}), (\ref{28}) as definitions of functions $M$ and $\kappa_A$ in terms of their initial values and functions $n_{AB}$. A more detailed discussion of these possibilities would be required if the Hamiltonian formalism based on (\ref{26}) proves to be useful.

In order to express the action (\ref{26}) in a form corresponding to the ADM action \cite{ADM} we proceed as follows. First we complete the metric by the term $-\epsilon dr^2$, where $\epsilon$ is a positive parameter. Then we find the ADM action related to the spacelike foliation by surfaces $u=$const. It contains terms which are singular at $\epsilon=0$. Since (\ref{26}) is not singular these terms have to be total derivatives. After removing them from the ADM action we take its limit when $\epsilon\rightarrow 0$. One can check directly that the resulting action differs from (\ref{26}) by a boundary term. Instead of the 3-dimensional scalar curvature present in the ADM action our action contains the scalar curvature of the metric $\tilde g_{AB}$ integrated over the coordinates $x^A$. This term can be removed due to the Gauss-Bonnet theorem. 
Finally, we obtain the following action integral  
\be 
I= \gamma\int d^4x\ \tilde \sigma [\dot {\tilde g}^{AB}\tilde p_{AB}-\tilde\lambda(\tilde p^A_{\ A,1}+\tilde p_{AB}\tilde p^{AB})-2\tilde \omega^A\tilde p_{AB}^{\ \ \ |B}-\frac{1}{2}\tilde g_{AB}\tilde\omega^A_{,1}\tilde\omega^B_{,1}].\label{30a}\ee
Here $x^1=r$, $\tilde \sigma=\sqrt{\det{\tilde g_{AB}}}$ and all operations on indices $A,B$, including covariant derivatives,  are defined by $\tilde g_{AB}$. Function $\tilde \lambda$ is related to $\tilde g_{00}$ via 
\be
\tilde g_{00}=\tilde \lambda+\tilde \omega_A\tilde\omega^A\label{26a}
\ee
and functions $\tilde p_{AB}$ are given  by
\be
\tilde p_{AB}=-\frac{1}{2}\tilde g_{AB,1}+(\ln\tilde \sigma)_{,1}\tilde g_{AB}\ .\label{31a}
\ee

In order to control behavior of fields at null infinity it is convenient to express the action (\ref{30a}) in terms of the unphysical metric $g=\Omega^2 \tilde g$
\be
g=du[(\lambda +\omega_A\omega^A)du-2d\Omega+2\omega_Adx^A]+g_{AB}dx^Adx^B\ .\label{29a}
\ee
Then one obtains
\be 
I= \gamma\int d^4x\ \tilde \sigma[\dot {g}^{AB}p_{AB}+\lambda(p^A_{\ A,1}-
p_{AB}p^{AB})-2\omega^Ap_{AB}^{\ \ \ |B}-\frac{1}{2}g_{AB}\omega^A_{,1}\omega^B_{,1}]\ ,\label{30}
\ee
where now $x^1=\Omega=\frac{1}{r}$, all operations on indices $A,B$ correspond to the metric $g_{AB}$, 
\be
p_{AB}=\frac{1}{2}g_{AB,1}-(\ln \sigma)_{,1}g_{AB}\label{31}
\ee
and $\sigma=\sqrt{\det{ g_{AB}}}$. 
 We notice that 
$\omega^A=\tilde\omega^A$, $\lambda =\Omega^2 \tilde \lambda$, $g_{AB} =\Omega^2 \tilde g_{AB}$ and $p_{AB}=\tilde p_{AB}-\Omega \tilde g_{AB}$. Conditions (\ref{4}) and (\ref{6}) are equivalent to 
\begin{equation}
\lambda=\Omega^{2}-2M\Omega ^3+O(\Omega^{4}),\label{32a}
\end{equation}
\begin{equation}
\omega _{A}=\Omega^{2}\psi_A+\Omega^3\kappa _A +O(\Omega^{4}).\label{31b}
\end{equation}
The asymptotic behavior of $g_{AB}$ is governed by (\ref{7}) and (\ref{8}).

Let us consider the action $I$ as a function of $\lambda$, $\omega^A$ and $g^{AB}$. The variable $\lambda$ appears in (\ref{30}) as a  Lagrange multiplier, in a similar way as the lapse function 
in the ADM formalism. The variation of $I$ with respect to $\lambda$ yields the equation 
\be
p^A_{\ A,1}-p_{AB}p^{AB}=0\label{32}
\ee
which is equivalent to the Einstein equation $\tilde G_{11}=0$. Equation (\ref{32}) is a constraint
 from the point of view of the evolution with respect to $u$ . In terms of $g_{\mu\nu}$ it 
reads
\be
\frac{1}{4}g_{AB,1}g^{AB}_{\ \ ,1}-(\ln \sigma)_{,11}=0.\label{33}
\ee
Equation (\ref{33}) reduces the number of independent degrees of freedom in $g_{AB}$ to two.

 The variation of (\ref{30}) with respect to $\omega^A$ leads to constraints 
\be
(\tilde \sigma g_{AB}\omega^B_{\ ,1})_{,1}-2\tilde \sigma p_{AB}^{\ \ \ |B}=0\label{34}
\ee
which are equivalent to the Einstein equations $\tilde G_{1A}=0$ (we keep $\tilde \sigma$ instead of $\sigma \Omega^{-2}$ in order to shorten notation).
It follows from (\ref{34}) that fields $\omega^A$ can be defined 
by a double integration of some functions of $g_{AB}$ and their  derivatives 
tangent to the surfaces of constant $u$. In this procedure the coefficients $\kappa_A$ appear as the integration constants (hence, condition (\ref{28}) can be imposed).

 The variation of (\ref{30}) with respect to $g^{AB}$ yields 
dynamical 
 equations equivalent to $\tilde G_{AB}=0$. They read
\begin{eqnarray}
&\dot g_{AB,1}+\frac{\tilde \sigma_{,1}}{2\tilde\sigma}\dot g_{AB}+\frac{\dot\sigma}{2\sigma}g_{AB,1}
+g^{CD}g_{C(A,1}(\omega_{B)|D}-\dot g_{B)D}-\frac{\lambda}{2}g_{B)D,1})-\frac{1}{2}(\omega^Cg_{AB,1})_{|C}\nonumber\\
&+\frac{1}{\tilde \sigma }(\frac{\lambda }{2}\tilde \sigma g_{AB,1}
-\tilde \sigma\omega_{(A|B)})_{,1}-\frac{1}{2}g_{AC}g_{BD}\omega^C_{\ ,1}\omega^D_{\ ,1}
+[\frac{\lambda}{2}S+\frac{1}{4}(\dot g_{CD}
-2\omega_{C|D})g^{CD}_{\ \ ,1}\nonumber\\
&-\frac{1}{4}g_{CD}\omega^C_{\ ,1}\omega^D_{\ ,1}+\omega^C(\ln \sigma )_{,1C}+\frac{1}{\tilde \sigma }(\tilde \sigma \omega^C_{\ |C})_{,1}
-\frac{1}{2\tilde \sigma}(\lambda \tilde \sigma)_{,11}
-\frac{1}{\tilde \sigma}\dot{\tilde \sigma } _{,1}-( \frac{\dot \sigma}{\sigma})_{,1}]g_{AB}=0,\label{35}
\end{eqnarray}
where $S$ denotes the l.h.s. of  (\ref{33}). 
 
Under conditions (\ref{4})-(\ref{8})
expanding equations (\ref{34}) and (\ref{35}) into powers of $\Omega$ implies relations (\ref{9}) and (\ref{17b}). It follows from them and from equations (\ref{27}) and (\ref{28}) that $\Omega^{-2}\tilde G_{0\mu} =0$ on $\I$ (equivalently, $\Omega^{-2}\tilde G^{1\mu} =0$ on $\I$ ). By virtue of the Bianchi identities these boundary conditions  assure that $\tilde G^{1\mu} =0$  everywhere.  Thus, the variational principle for (\ref{26}) yields a complete set of the Einstein equations under assumptions (\ref{27}) and (\ref{28}). 

The Einstein equation $\tilde G^{01}=0$ follows from (\ref{33})-(\ref{35}) independently of the boundary equations (\ref{27}) and (\ref{28}). It takes the form
\be
\frac{1}{2}g_{AB}\omega^A_{,1}\omega^B_{,1}-(\omega^A,_1)_{|A}+\tilde\sigma^{-1}(2\tilde\sigma\omega^A_{\ |A}-2\tilde\sigma,_0-\lambda\tilde\sigma,_1),_1-R^{(2)}=0\ ,\label{35a}
\ee
where $R^{(2)}$ is the Ricci scalar of the metric $g_{AB}$. Equation  
(\ref{35a}) can be used to eliminate one of equations  (\ref{35}) and to express $\lambda$ in terms of other quantities (eventually one can reduce  (\ref{35}) to two equations for two components of $g_{AB}$).
It is also important for the Hamiltonian description in the next section.

\section{Hamiltonian}

\null
In this section we consider a canonical formulation based on the Lagrangian density  
\be 
\mathcal {L}=\gamma\tilde \sigma [\dot {g}^{AB}p_{AB}+\lambda(p^A_{\ A,1}-
p_{AB}p^{AB})-2\omega^Ap_{AB}^{\ \ \ |B}-\frac{1}{2}g_{AB}\omega^A_{,1}\omega^B_{,1}]\label{36a}
\ee
corresponding to the action (\ref{30}). 
This formulation differs substantially from other approaches (see eg. \cite{J} and references therein) since in metric (\ref{3}) there is no more freedom of the lapse and shift functions. Due to this fact a structure and interpretation of constraints is violated. Probably the biggest advantage of this formalism is a relative simplicity of the Lagrangian and the corresponding Hamiltonian. Also results of the next section can be interesting . Nevertheless, this section has rather purely informative character without definite conclusions about applications to the light cone quantization or characteristic initial value problem.

It follows from (\ref{36a}) that $\lambda$ is a Lagrange multiplier. For this reason we take 
 the space of fields $(g^{AB},\omega^A)$ as the configuration space.  
Elements of the phase space are given by $(g^{AB},\omega^A, \pi_{AB},\Pi_{A})$, where $\pi_{AB}$ and $\Pi_{A}$ are the  momenta conjugate, respectively, to $g^{AB}$ and $\omega^A$. Since the Lagrangian density $\mathcal L$  is linear in $\dot g^{AB}$  the inverse Legendre transformation is not well defined.  Following the standard procedure of Dirac \cite{D}\cite{H} we treat equations
\be
\pi_{AB}-\gamma\tilde \sigma p_{AB}=0\label{36}
\ee 
as constraints which should follow from a total Hamiltonian. Apart from that, since $\mathcal L$ contains no time derivative of $\omega^{A}$ we must add further constraints
\be
\Pi_A=0.\label{37}
\ee 
Constraints (\ref{36}), (\ref{37}) together with (\ref{33}) are primary constraints of the theory. 
 Taking them into account leads to the following completion of the canonical Hamiltonian density $\pi_{AB}\dot g^{AB}-\mathcal {L}$
\be 
\mathcal {H}=\gamma\tilde \sigma [2\omega^Ap_{AB}^{\ \ \ |B}+\frac{1}{2}g_{AB}\omega^A_{,1}\omega^B_{,1}-\lambda S]+\mu^{AB}\phi_{AB}+\nu^A\Pi_A ,\label{38}
\ee
where $\mu^{AB}$, $\nu^A$ are Lagrange multipliers and $\phi_{AB}$ denotes
the l.h.s. of constraint (\ref{36}).
The Hamiltonian functional is defined as $H=\int d^{3}x\mathcal{H}$.

The next step of Dirac's algorithm is to assure a proper propagation of all constraints. From
\be
\{\phi_{AB},H\}=0 \label{39}
\ee
we get complicated equations which do not involve time derivatives and can be treated as  conditions imposed on the multipliers $\mu^{AB}$ and $\lambda$. By virtue of the Hamilton equations 
\be
\dot g^{AB}=\frac{\delta H}{\delta\pi_{AB}}=\mu^{AB}\ ,\ \ \dot \omega^A=\frac{\delta H}{\delta \Pi_A}=\nu^A\label{40}
\ee
equations (\ref{39}) become equivalent to (\ref{35}) (using this fact one can easily reconstruct an explicit form of (\ref{39})). Thus, the dynamical equations    (\ref{35}) are  encoded in the conditions on the Lagrange multipliers. 

The propagation of the constraint (\ref{33}) gives another condition on $\mu^{AB}$
\begin{equation}
\mu^{AB}_{\ \ ,1}g_{AB,1}-\mu_{AB,1}g^{AB}_{\ \ ,1}+2\mu^A_{\ A,11}=0\ ,\label{40a}
\end{equation}
whereas constraint (\ref{37}) leads to equation (\ref{34}). Since (\ref{34}) does not contain any Lagrange multiplier it is a secondary constraint of the theory. 
Its  time derivative  yields equations 
\begin{eqnarray}
&\frac{1}{\tilde \sigma}[\tilde \sigma g_{AB}\nu^B_{,1}-\tilde \sigma \omega^B_{,1}(\mu_{AB}+\frac{1}{2}\mu g_{AB})]_{,1}+(\mu_{AB,1}-\mu_B^{\ C}g_{AC,1}+\frac{1}{2}\mu g_{AB,1})^{|B}\nonumber\\
&-\frac{1}{2}g_{BC,1}\mu^{BC}_{\phantom{BC}|A}-\mu_{,1A}-\mu(\ln \sigma)_{,1A}=0, \label{43}
\end{eqnarray} 
(here $\mu=\mu^A_A$)  
which can be considered as conditions on the multipliers  $\nu^{A}$. 

 Thus, we end up with conditions (\ref{39}), (\ref{40a}), (\ref{43}) on the Lagrange multipliers and with constraints (\ref{33}), (\ref{34}), (\ref{36}), (\ref{37}) on the fields $g^{AB},\omega^A$ and their momenta $\pi_{AB},\Pi_{A}$. All these constraints are second class in Dirac's terminology (the lack of first class constraints is related to fixing coordinates in metric (\ref{3})). The non vanishing Poisson brackets of the constraints are given by 
\begin{eqnarray}
&\gamma^{-1}\{\phi_{AB},\phi(\alpha^{CD})\}=\tilde \sigma \alpha_{AB,1}+\frac{1}{2}\alpha_{AB}\tilde \sigma_{,1}+\frac{1}{4}\alpha \tilde \sigma g_{AB,1}-\tilde \sigma \alpha^C_{\phantom{I}(A}g_{B)C,1}\nonumber\\
&-[\frac{1}{4}\tilde \sigma \alpha^{CD}g_{CD,1}+\tilde \sigma\alpha_{,1}+\frac{1}{2}\alpha\tilde \sigma _{,1}]g_{AB} \label{44}\\ 
&\{\phi_{AB},S(\beta)\}=-\frac{1}{2}g_{AB}\beta_{,11}+\frac{1}{2}\beta_{,1}  g_{AB,1}+\frac{1}{2} \beta (g_{AB,11} -g^{CD}g_{AC,1}g_{BD,1})\label{45}\\
&\{\psi_A,\phi(\alpha^{BC})\}=-\frac{1}{\tilde \sigma}[\tilde \sigma \omega^B_{,1}(\alpha_{AB}+\frac{1}{2}\alpha g_{AB})]_{,1}+(\alpha_{AB,1}-\alpha_B^{\ C}g_{AC,1}+\frac{1}{2}\alpha g_{AB,1})^{|B}\nonumber\\
&-\frac{1}{2}g_{BC,1}\alpha^{BC}_{\phantom{BC}|A}-\alpha_{,1A}-\alpha(\ln \sigma)_{,1A}\label{46}\\
&\{\Pi_A,\psi(\gamma^B)\}=-(\tilde \sigma g_{AB}\gamma^B_{\ ,1})_{,1}.\label{47}
\end{eqnarray}
Here $\psi_A$ denotes the l.h.s. of (\ref{34}) and $\alpha^{AB}$, $\beta$, $\gamma^A$ are smooth smearing functions (e.g. $\psi(\gamma^A)=\int d^3x \psi_A\gamma^A$ and $\alpha=\alpha^A_{\ A}$).  

An alternative possibility of the Hamiltonian formulation is to treat the variables $\omega ^A$ in the Lagrangian (\ref{35a}) as functions of $g_{AB}$ defined by equations (\ref{34}) (the resulting expressions for $\omega ^A$ are still simpler than those for $\nu^{A}$ in the previous formulation). The corresponding Euler-Lagrange equations remain equivalent to the Einstein equations.  Now the phase space consists of elements $(g^{AB}, \pi_{AB})$. The total Hamiltonian density is given by (\ref{38}) with $\nu^A=0$. Equations (\ref{33}) and (\ref{36}) are the primary constraints. Their
 propagation  yields  equations (\ref{39}) and (\ref{40a}) which define the Lagrange multipliers $\lambda$ and $\mu^{AB}$. In this formulation the number of constraints and Lagrange multipliers is substantially reduced but still the Hamiltonian picture remains quite complicated.

\section{Energy}

 We notice that the Hamiltonian density (\ref{38}) is not a sum of constraints, unlike in the ADM calculus. When the field equations  are satisfied then $\mathcal {H}$ takes the value
\be 
\mathcal {H}=\gamma(\tilde\sigma\omega_A\omega^A,_1),_1-\frac{\gamma}{2}\tilde\sigma g_{AB}\omega^A_{,1}\omega^B_{,1}\ .\label{48}
\ee
(Thus, it may be appealing to add the term  $-\gamma(\tilde\sigma\omega_A\omega^A,_1),_1$ in (\ref{38}) to obtain Hamiltonian
which is nonnegative for fields satisfying the Einstein equations.) 
Due to identity (\ref{35a}) 'on shell' value of $\mathcal {H}$ can be expressed in the following way
\be 
\mathcal {H}=\gamma[f,_1+2(\tilde\sigma\omega^A_{\ |A}),_1-\tilde\sigma(\omega^A,_1)_{|A}-2\Omega^{-2}\sigma_s-\tilde
\sigma R^{(2)}] ,\label{50}
\ee
where $\sigma_s=\sqrt{\det{s_{AB}}}$ and 
\be
f=\tilde\sigma\omega_A\omega^A,_1-2\tilde\sigma,_0-\lambda\tilde\sigma,_1-2\Omega^{-1}\sigma_s\ .\label{51}
\ee
Integrating (\ref{50}) over a part of the null hypersurface $u=$const contained between the scri $\I$ and the surface $\Omega=$const yields 
\be
\int _0^{\Omega}\mathcal {H}=E+\gamma\int_{\Sigma} f(\Omega)\ ,\label{52}
\ee
where $E$ is the energy (\ref{10a}) related to the section $u=$const of $\I$ and the integral on the r.h.s. is taken over the coordinates $x^A$ 
(note that the integral of the last two terms in (\ref{50}) vanish due to the Gauss-Bonnet theorem). Thus, as in standard Hamiltonian formulations \cite{KT}, $H$ becomes a boundary integral for fields satisfying the Einstein equations.

Assume that the surface $u=$const has topology of a light cone in Minkowski space with vertex at $r=\Omega^{-1}=0$. If $f=0$ at $r=0$  then the integral of $\mathcal {H}$ over the cone coincide with the energy (\ref{10a}). Moreover, it follows from  (\ref{48}) and (\ref{52}) that 
\be
E\geq \int_{\Sigma}\tilde f(\Omega)\label{53}
\ee
where 
\be
\tilde f=2\tilde\sigma,_0+\lambda\tilde\sigma,_1+2\Omega^{-1}\sigma_s\ .
\label{54}
\ee
If $\tilde f=0$ at $r=0$, or e.g. $\tilde f\geq 0$ for some $\Omega$, then the total energy has to be nonnegative (in accordance with known results \cite{LV}\cite{IN}). For instance, the functions $f$ and $\tilde f$ vanish at $r=0$  if $\tilde g_{AB}$ is of order $r^2$ and $\tilde\omega_A$ of order $r$ near the vertex. These assumptions are difficult to control in  the framework of  the Bondi-Sachs approach. Perhaps spacetime can be sliced by 'good cones' due to a matter source and then the BMS group is reduced to the Poincare group. Note that inequality (\ref{53}) and its consequences are independent of our choice of the Hamiltonian.
 
\section{Concluding remarks}

We transformed the Hilbert action for the Bondi-Sachs metrics to the forms (\ref{30a}) and (\ref{30}) similar to the ADM action. The variational principle for the action (\ref{30}) yields equations (\ref{33})-(\ref{35}) which are equivalent to the Einstein vacuum equations provided the  r-independent equations (\ref{27}), (\ref{28}) are satisfied. Equations  (\ref{33}), (\ref{34}) are constraints from the point of view of the evolution with respect to the Bondi time $u$. They can be solved in terms of integrals over $\Omega$ (or $r$).

We analyzed briefly the corresponding  Hamiltonian formulation. The Dirac method for constrained systems leads to the Hamiltonian (\ref{38}) (or its truncation with $\nu^A=0$) with the Lagrange multipliers which carry the main information about dynamics of the system.  The Lagrange multipliers  are defined implicitly via a complicated system of differential equations. There are six second class constraints and no first class constraints.

The Hamiltonian does not vanish when the Einstein equations are satisfied however it becomes a boundary integral. One can modify the  Hamiltonian density in order to obtain an expression which takes nonnegative 'on shell' values. Under some assumptions about a structure of surfaces $u=$const the total energy is  also nonnegative in accordance with known results.

\bigskip
\noindent {\bf Acknowledgements}. We greatly acknowledge a suggestion of one of referees that 'on shell' Hamiltonian has to be a boundary term.

This work was partially supported by the Polish  Committee for Scientific Research (grant 2 PO3B 12724). JK acknowledges the support of European Commission through the project ``EQUIP'' within the framework of the IST-program.

\end{document}